\def\gsimeq
{\hbox{\raise0.5ex\hbox{$>\lower1.06ex\hbox{$\kern-1.07em{\sim}$}$}}}
\def\lsimeq
{\hbox{\raise0.5ex\hbox{$<\lower1.06ex\hbox{$\kern-1.07em{\sim}$}$}}}
\def\pn{\par\noindent}

\def\bs{\bigskip\pn}

\documentclass[cup5b]{caps.manyauthors2}

\usepackage{graphicx,psfig}
\usepackage{amssymb}
\usepackage{ociwsymp1e}
\HeadText{M. Cappi et al.}

\begin{document}

\pagenumbering{arabic}

\author[]{
M. CAPPI$^{1}$, G. DI COCCO$^{1}$, F. PANESSA$^{1,7}$, L. BASSANI$^{1}$, E. 
CAROLI$^{1}$, \cr M. DADINA$^{1}$, A. COMASTRI$^{2}$, R. DELLA CECA$^{3}$, 
A. V. FILIPPENKO$^{4}$, L. FOSCHINI$^{1}$, \cr F. GIANOTTI$^{1}$, L. C. 
HO$^{5}$, M. MAKISHIMA$^{6}$, G. MALAGUTI$^{1}$, J. S. MULCHAEY$^{5}$, \cr
G. G. C. PALUMBO$^{7}$, E. PICONCELLI$^{1,7}$, W. L. W. SARGENT$^{8}$, J. 
STEPHEN$^{1}$, M. TRIFOGLIO$^{1}$, \cr K. A. WEAVER$^{9}$ and  G. 
ZAMORANI$^{2}$ \\
(1) Istituto IASF-CNR, Sezione di Bologna, Italy; 
(2) INAF-Bologna, Italy;
(3) INAF-Milano, Italy;\\
(4) UC-Berkeley, USA;
(5) Carnegie Observatories, USA;
(6) University of Tokyo, Japan;\\
(7) Astronomy Department of Bologna, Italy;
(8) Caltech, USA;
(9) NASA/GSFC, USA }

\chapter{{\it XMM-Newton}\ Survey of a \\ Distance-limited 
Sample of Seyfert Galaxies}

\bs
\bs

\begin{abstract}

An unbiased estimate of the average intrinsic X-ray properties and column density distribution 
of Seyfert galaxies in the local Universe is crucial to validate unified models of active galactic 
nuclei (AGNs) and to synthesis models for the X-ray background. 
We present here preliminary results obtained from an
on-going $XMM-Newton$ study ($\sim$250 ks awarded in the EPIC GT) on a well-defined, 
statistically complete, and significant sample of nearby Seyfert galaxies.

\end{abstract}

\section{Introduction}

It has become clear, after $ASCA$ and $BeppoSAX$ results, that X-ray observations of Seyfert galaxies 
are key to verifying the predictions and, thus, the validity of unified models of active galactic 
nuclei (AGNs). They are clearly also crucial to test the applicability of standard 
accretion disk theories down to the lowest (L$_{B}$=10$^{8}$-10$^{11}$ L$_{\odot}$) 
nuclear luminosities. 


Hard X-ray samples of nearby Seyfert galaxies available to date (e.g. from 
$GINGA$, $ASCA$ and $BeppoSAX$; Smith \& Done 1996, Turner et al., 1998, Bassani et al. 1999) 
have been necessarily biased towards the most X-ray luminous, and less absorbed AGNs. However, 
studies by Maiolino et al. (1998) and Risaliti, Maiolino and Salvati (1999) 
suggest the existence of a large fraction ($\gsimeq$50-60\%) of highly obscured AGNs in the local Universe. 
Other recent studies by Ho et al. (2001) with $Chandra$
have made possible, for the first time, the detection of point-like nuclear sources in a large sample 
of nearby galaxies at unprecedented low luminosities (down to less than 10$^{38}$ erg/s), and 
such studies seem to suggest that the standard unified model for Seyfert galaxies (Antonucci 1993) may not 
hold down to these very low luminosities.

With the aim of sorting out the questions raised by the above observations, here new results obtained 
with the EPIC CCDs on-board $XMM-Newton$ are presented. We are now completing a survey of all known 
northern Seyfert galaxies with D$<$22 Mpc using the Revised-Ames Catalog 
of Bright Galaxies (Sandage \& Tamman 1981) spectroscopically classified by 
Ho, Filippenko \& Sargent (1997). The final sample contains 28 galaxies (see Table 1 in Cappi et al., 2002)
and is the deepest and most complete local sample of Seyfert galaxies in the optical energy band. 
Compared to previous studies, the strenght of using $XMM-Newton$ rests on two main facts: 
its high-throughput (especially at E$>$2 keV) allows to search spectral 
components with absorption columns up to $N_{\rm H}$ $\sim$ 10$^{22}$-10$^{24}$ cm$^{-2}$, and 
its spatial resolution (HPR$\simeq$7 arcsec) allows to minimize any strong contamination 
from off-nuclear sources to the soft (E$<$ 2 keV) and/or hard (E$>$2 keV) energy band.

\section{Spectral Analysis: Preliminary Results}

About 250 ksec of EPIC Guaranteed Time (Di Cocco et al., 2000) have been given to our team 
to perform this survey.
To date, 20 out of 28 Seyfert galaxies have been observed, and the data analyzed. Exposures 
of 5, 10, 15 and 20 ks, were requested in order to obtain $\sim$1000 source counts per galaxy.
Spectra were fitted using data from the 3 CCD detectors (MOS1, MOS2 and PN) simultaneously.
Data were fitted between 0.3-10 keV for MOS1/2 and between 0.5-10 keV for the PN.

Every spectrum has been fitted using a set of standard models: a single power-law model with 
Galactic absorption plus either
i) free absorption, or ii) a warm-absorber, or iii) a single soft-scattered component, or iv) 
a single soft thermal component, or v) a scattered plus thermal soft component.
We then added a narrow FeK emission line if detected.
Six examples of such spectra are shown in Figures 1.1, 1.2 and 1.3 to illustrate the data available
 and some of our early results. 
Best-fit spectra and model parameters will be reported elsewhere (Cappi et al., in preparation).

Figure 1.1 shows two examples of Seyfert 1.5 galaxies (NGC5273 and NGC3227) that do have 
significant ionized absorption (with ionized columns equivalent to $\sim$ 1 and 7 $\times$ 10$^{22}$ cm$^{-2}$, 
respectively). Figure 1.2 shows two examples of Seyfert 2 galaxies (NGC4725 and 
NGC4168) that do not have significant absorption in excess of the Galactic value. 
Figure 1.3 (left) illustrates a case of an
underluminous Seyfert 2 galaxy (L$_{\rm 2-10 keV}$ $\sim$ 5 $\times$ 10$^{38}$ erg/s, to be compared 
to an expected value, based on the observed L$_{H\alpha}$, that is $\sim$ 40 times larger).
Figure 1.3 (right) shows the single clear case (NGC3079) yet in our sample of a Compton thick source, 
that is evidenced by the strong FeK line (with EW$\sim$1.9$\pm$0.5 keV).  

In addition to the high-quality $XMM-Newton$ images available, we also 
checked the $Chandra$ snapshot observations (available for 12 sources out of the 20 analyzed yet) 
to check for possible bright off-nuclear point sources that may have contaminated the 
$XMM-Newton's$ spectra. We estimate that off-nuclear point-sources and/or diffuse emission may 
contaminate both the soft and hard emission in less than $\sim$ 1/5$^{\rm th}$ of the sources, 
while up to $\sim$ 1/3$^{\rm rd}$ of the sources could be contaminated in the soft 
band only. Further work is under way to quantify and take this effect into account in the scientific analysis.

\section{Conclusions}

Results so far obtained can be summarized as follows:

i) Three out of the five Seyfert 1/1.5 present in the sample do show heavy ($\sim$10$^{22}$ cm$^{-2}$) 
ionized absorption in X-rays, in agreement with what found also in higher-luminosity 
type 1 AGNs (e.g. Reynolds 1997).

ii) Most Seyfert 1.9$/$2 are underluminous in X-rays with respect to what expected from 
their H$_{\alpha}$ luminosities (Cappi et al., in preparation), in agreement with what was found 
by Ho et al. (2001) and Terashima et al. (2000).

iii) Most Seyfert 1.9/2 do not show neither strong absorption, nor strong FeK lines. In some sources, 
the low absorption could be due to the extra contribution in soft X-rays from nearby off-nuclear sources. 
We estimate however that this effect is not relevant in the majority of the cases studied so far.
The possibility that such sources might be Compton-thick appears also unlikely given the 
absence of a strong FeK line (except for NGC3079).
If confirmed, these results appear to be in contrast with the simplest version 
of unified models.

Correlation of X-ray data  with emission at other wavelengths and statistical properties of 
the whole sample are under-way.

\begin{figure}[!]
\hspace{-1cm}
\parbox{6cm}{
\psfig{file=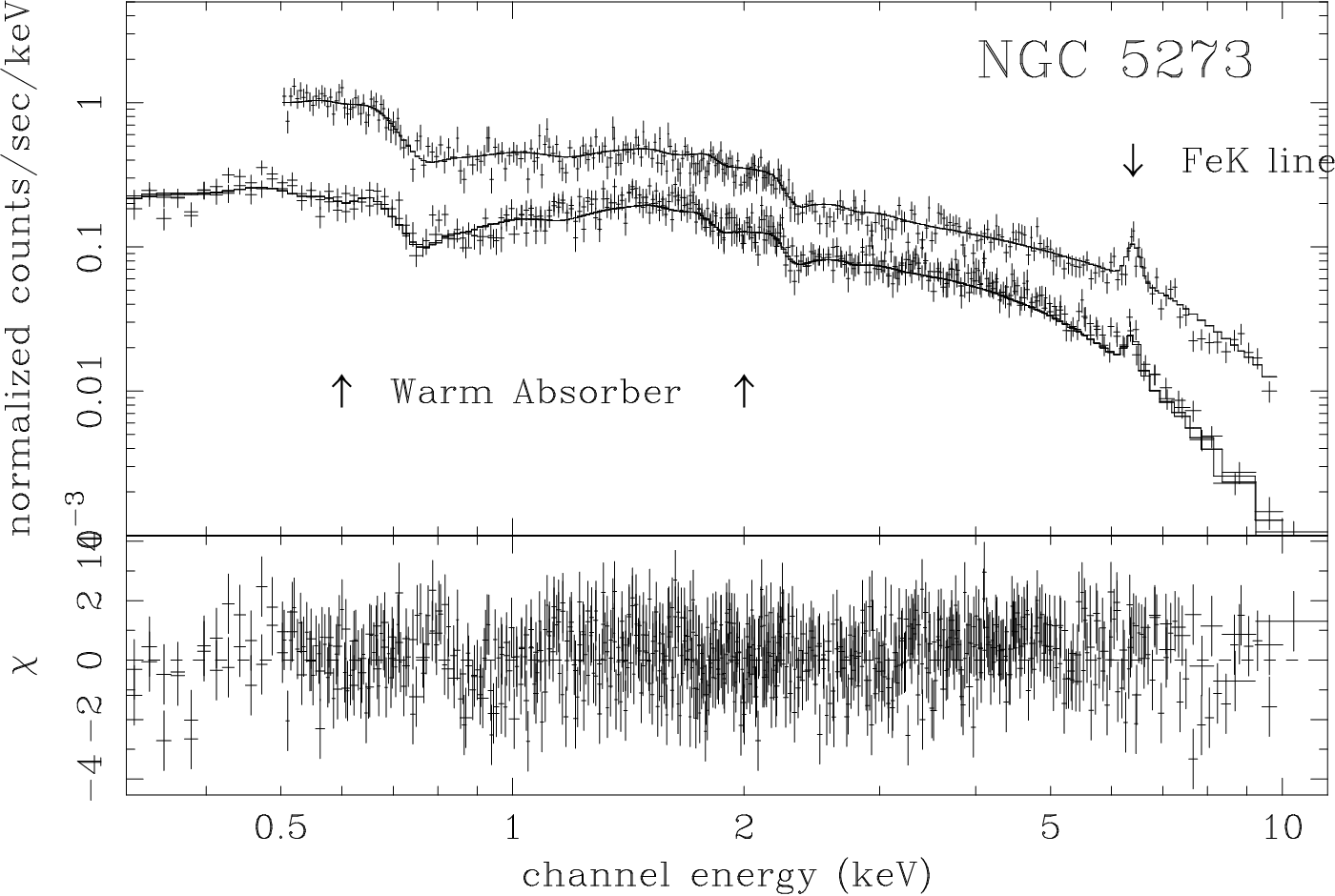,width=6cm,height=5.5cm,angle=-90}}
\hspace{0cm} \
\parbox{6cm}{
\psfig{file=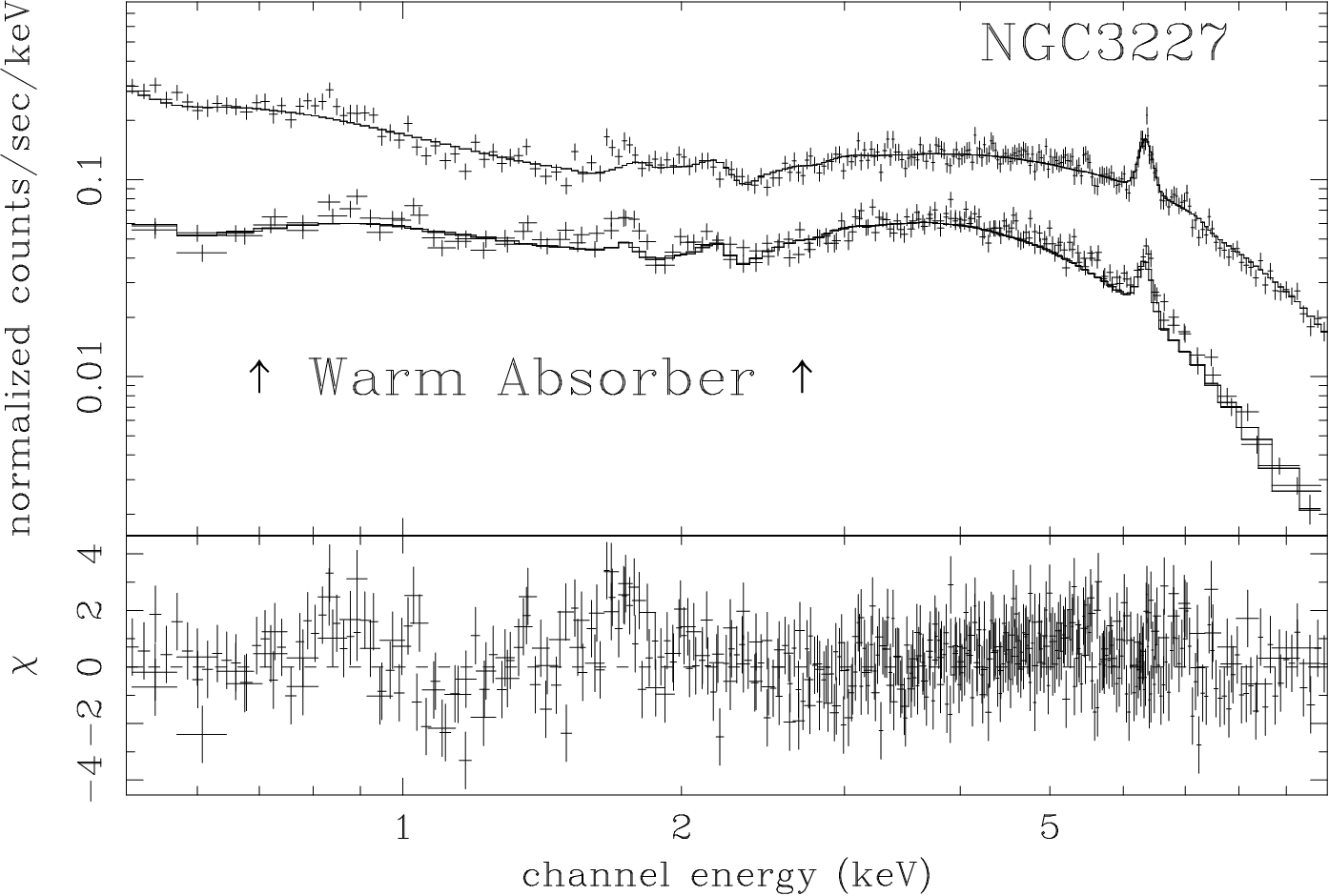,width=6cm,height=5.5cm,angle=-90}}
\footnotesize{Fig. 1: MOS and PN spectra of the two Seyfert 1.5s: NGC5273 (left) and NGC3227 (right). 
The spectra of these type-1 sources illustrate the effects of a strong warm absorber in these sources.}
\end{figure}

\begin{figure}[!]
\hspace{-1cm}
\parbox{6cm}{
\psfig{file=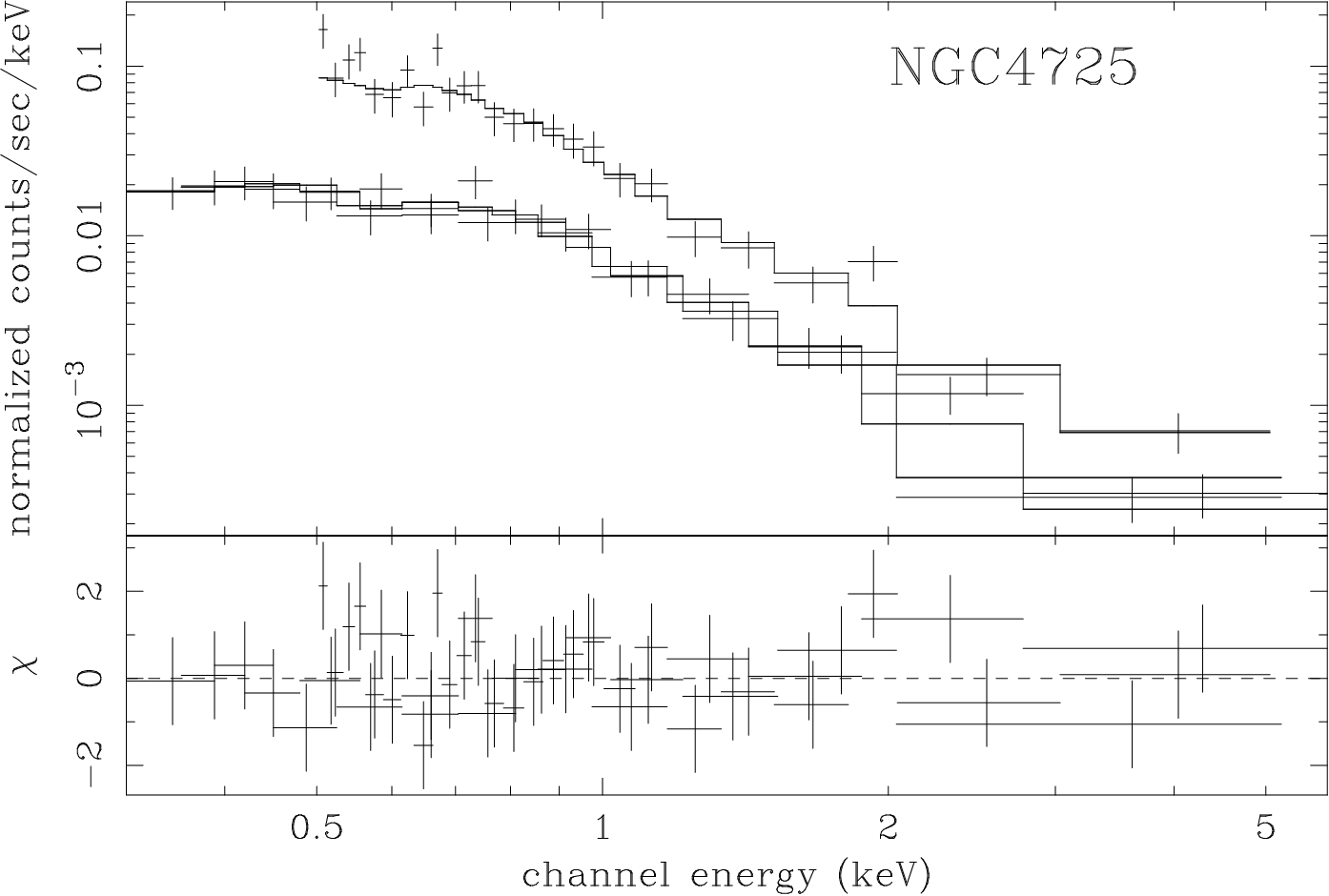,width=6cm,height=5.5cm,angle=-90}}
\hspace{0cm} \
\parbox{6cm}{
\psfig{file=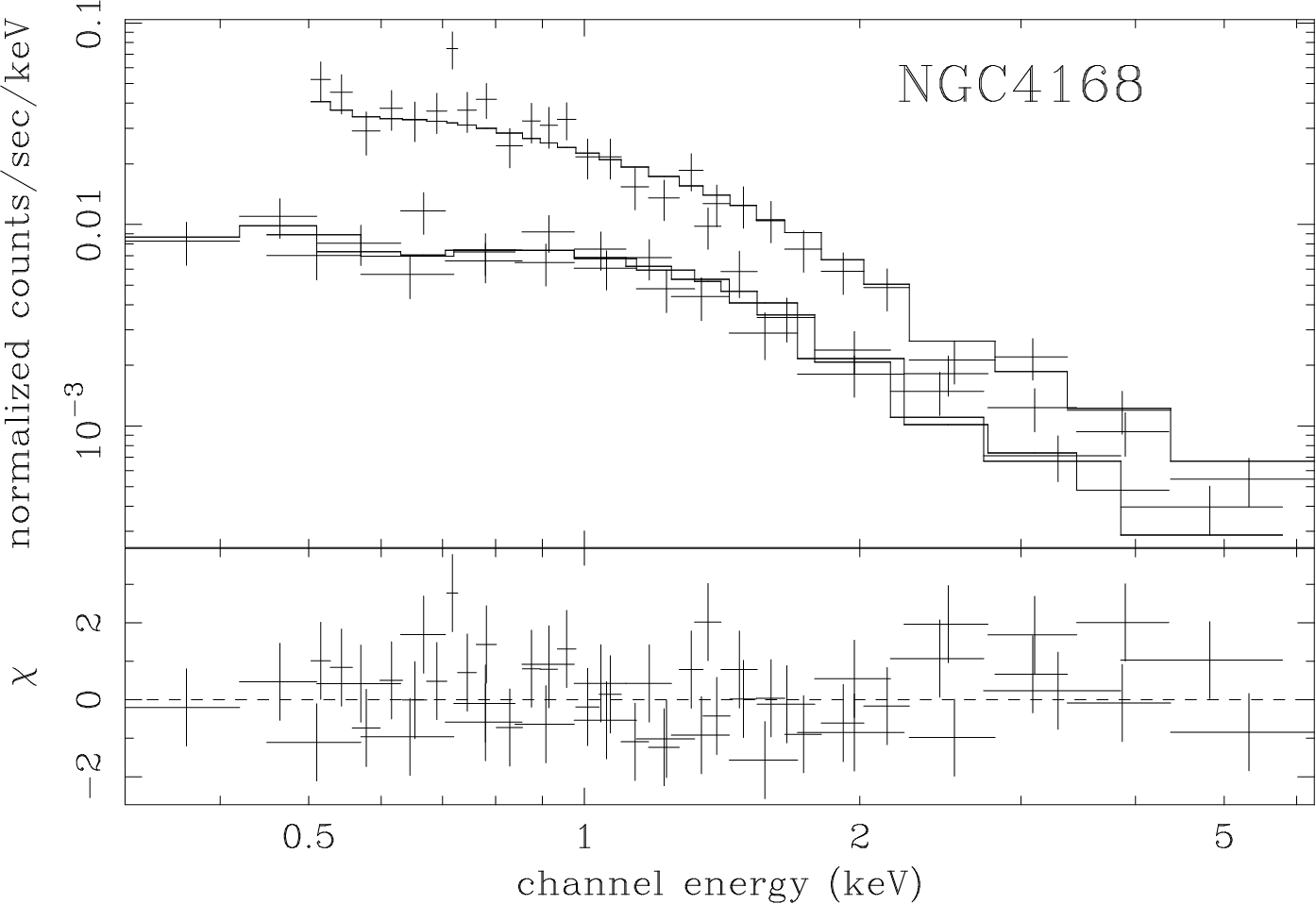,width=6cm,height=5.5cm,angle=-90}}
\footnotesize{Fig. 2: MOS and PN spectra of the two Seyfert 2s: NGC4725 (left) and NGC4168 (right).
The spectra of these type-2 sources illustrate well the lack of any absorption in these sources.}
\end{figure}

\begin{figure}[!]
\hspace{-1cm}
\parbox{6cm}{
\psfig{file=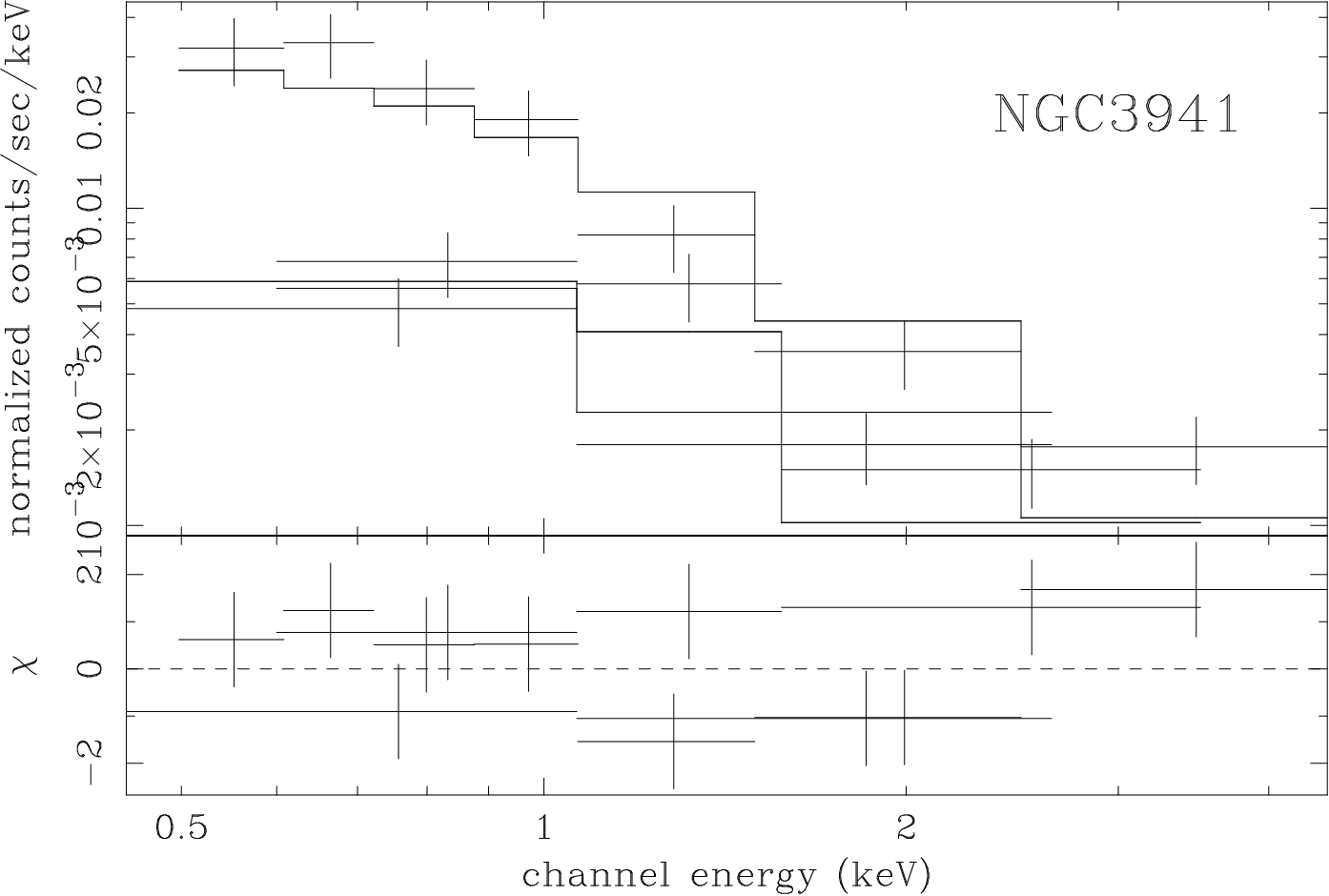,width=6cm,height=5.5cm,angle=-90}}
\hspace{0cm} \
\parbox{6cm}{
\psfig{file=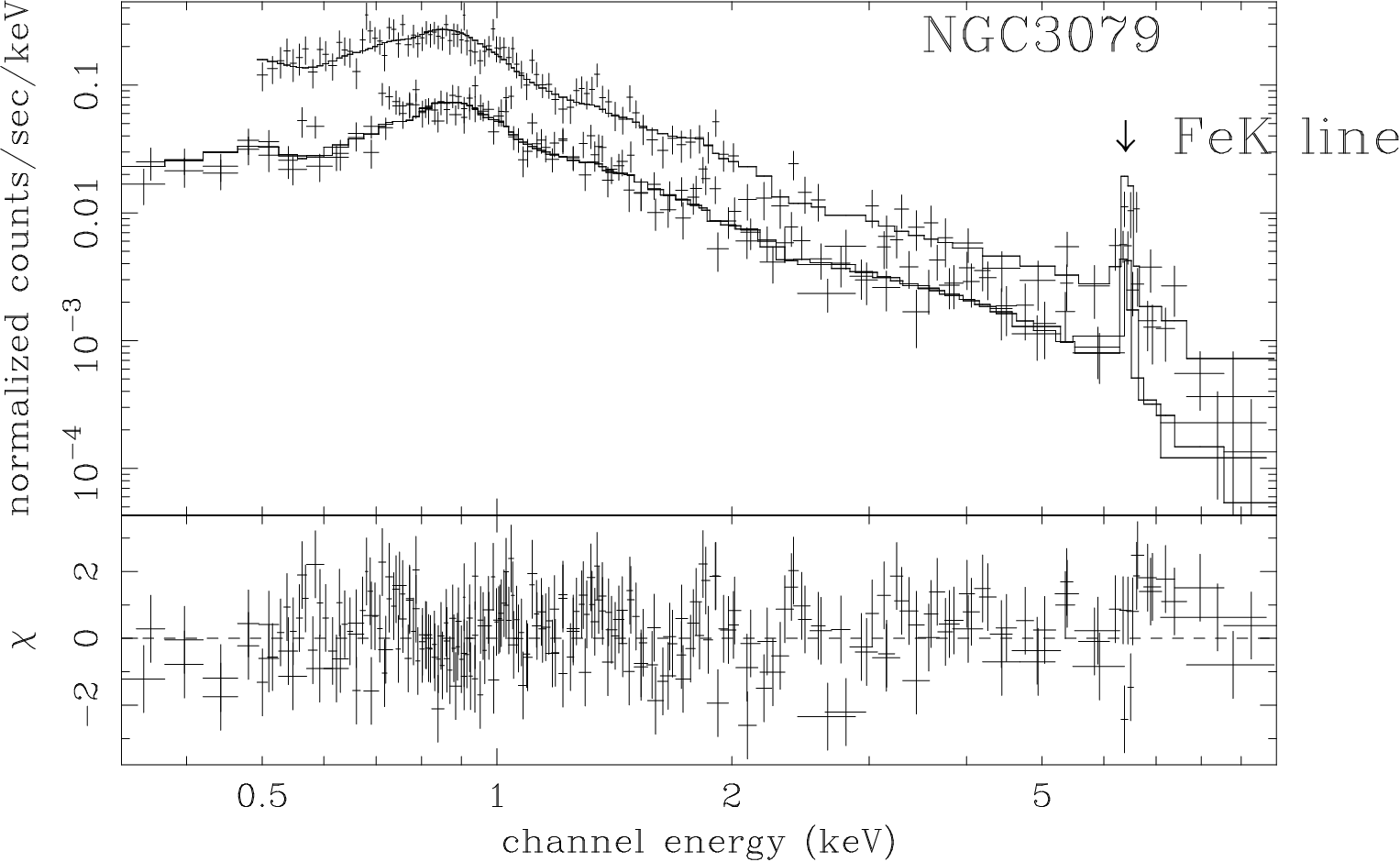,width=6cm,height=5.5cm,angle=-90}}
\footnotesize{Fig. 3: MOS and PN spectra of the two Seyfert 2s: NGC3941 (left) and NGC3079 (right).
NGC3941 illustrates the case for the source with lowest luminosity of our sample (and very 
underluminous with what expected from its H$_{\alpha}$ luminosity) and NGC3079 the case 
for a Compton-thick source (see the strong FeK line detected).}
\end{figure}

\end{document}